\documentclass{aa}  

\usepackage{graphicx}
\usepackage{lscape}
\usepackage{txfonts}
\begin{document}
   \title{On the circum(sub)stellar environment of brown dwarfs in Taurus
   \thanks{Based on observations made at ESO, CFHT, 2MASS, \& {\em Spitzer}.}}

   \subtitle{}

   \author{S.~Guieu
             \inst{1}
          \and
          C.~Pinte\inst{1}
           \and
          J.-L.~Monin\inst{1,2}
            \and
          F.~M\'enard\inst{1}
            \and
          M.~Fukagawa\inst{3}
            \and
          D.~L.~Padgett\inst{3}
          \and A.~Noriega-Crespo\inst{3}
          \and S.~J.~Carey\inst{3}
          \and L.~M.~Rebull\inst{3}
          \and T.~Huard\inst{4}
          \and M.~Guedel\inst{5}
          }

   \offprints{S. Guieu}
   \institute{Laboratoire d'Astrophysique de Grenoble,  BP 53, 38041 Grenoble, France 
              \email{Sylvain.Guieu@obs.ujf-grenoble.fr}
               \and Institut Universitaire de France 
         \and Spitzer Science Center, Caltech, Pasadena, CA 91125, USA 
             \and Harvard-Smithsonian Center for Astrophysics, 60 Garden Street, Cambridge, MA 02138, USA
             \and Paul Scherrer Institut, W\"urenlingen and Villigen, CH-5232 Villigen PSI, Switzerland
             }

   \date{Received 31 July 2006 ; Accepted 18 December 2006}

 
  \abstract
  {}  
   {We want to investigate whether brown dwarfs (BDs) form like stars
or are ejected embryos. We study the presence of disks around BDs in
the Taurus cloud, and discuss implications for substellar formation
models.}
   {We use photometric measurements from the visible to the far
infrared to determine the spectral energy distributions (SEDs) of
Taurus BDs.}
     {We use Spitzer color indices, H$\alpha$ as an accretion 
indicator, and models fit to the SEDs in order to estimate
physical parameters of the disks around these BDs.  We study the
spatial distribution of BDs with and without disks across the
Taurus aggregates, and we find that BDs with and without disks
are not distributed regularly across the Taurus cloud.}
  {We find that $48\%\pm14\%$ of Taurus BDs have a circumstellar disk
signature, a ratio similar  to recent results from previous authors 
 in other regions. We fit the SEDs and find that none of the disks
around BDs in Taurus can be fitted convincingly with a flaring index
$\beta = 0$, 
indicating that heating by the central object is efficient and that
the disks we observe retain a significant amount of gas.  We find that
BDs with disks are proportionally more numerous in the northern Taurus
filament, possibly the youngest filament.  We do not find such a clear
segregation for classical T~Tauri stars (CTTS) and weak-lined
T~Tauri stars (WTTS), suggesting that, in addition to the
effects of evolution, any segregation effects could be related to the
mass of the object.  A by-product of our study is to propose a
recalibration of the Barrado y  Navascu\'es \& Mart\'in (2003)
accretion limit in the substellar domain. The global shape of the
limit fits our data points if it is raised by a factor 1.25-1.30.
}


   \maketitle
%

\section{Introduction}

In recent years, a large number of brown dwarfs (BDs) have been
detected in star forming regions, opening the opportunity to study the
stellar formation process and the corresponding IMF deep into the
substellar domain, even down to the planetary mass regime (Chauvin et
al.\ 2005; Luhman et al.\ 2005).  Two main classes of models have been
proposed for the formation of substellar objects. In the {\sl standard
formation scenario}, BDs form like stars, through (turbulent)
gravitational collapse and fragmentation of very low mass cores,
followed by subsequent disk accretion. In the {\sl ejection model},
BDs are stellar embryos ejected from their parent core either early in
their evolution from dynamically unstable multiple protostellar
systems (Reipurth \& Clarke 2001), or through secular dynamical decay
in dense embedded clusters (Sterzik \& Durisen 2003; Kroupa \& Bouvier
2003a). These two models are not mutually exclusive; other mechanisms
are also discussed in the literature (e.g., Whitworth \& Goodwin 2005,
Whitworth et al.\ 2006).

The Taurus region has been extensively studied for star formation. It
is young (1-5 Myr), so the dynamical effects remain limited. It
extends over a large region, so it can be studied for large spatial
distribution effects; there are no bright stars to irradiate and
disturb the stellar surroundings.  Our study is based on a sample of 33 BDs in the
Taurus cloud as presented in Guieu et al.\ (2006) and references
therein.  With such a number of objects at hand, we can now begin
statistical studies. With the aim of studying the proportion of Taurus
BDs that harbor an accretion disk, we have combined Guieu et al.\
(2006) optical photometry with $JHK_s$ 2MASS data and recent {\it
Spitzer} 3.6 to $70\,\mu$m data to determine the spectral energy
distributions (SEDs) of BDs in the Taurus cloud. 
%

Numerous studies have attempted to distinguish between the two
principal models of BD formation by examining the circumstellar disks
of young brown dwarfs.  For instance, there is now ample evidence
that, like their more massive counterparts, Taurus BDs experience a
T~Tauri phase.  Broad asymmetric H$\alpha$ emission profiles
characteristic of accretion have been found; see, e.g., Jayawardhana
et al.\ (2003) and Muzerolle et al.\ (2005). $L$ band excesses have
been detected in Taurus substellar sources, indicating a disk
frequency $\approx 50\,$\% (Liu et al.\ 2003). Recently, Whelan et
al.\ (2005) have detected an outflow from a BD by spectroastrometry. 
It has often been argued that the presence of accretion and/or outflow
activity in BDs is evidence that brown dwarfs form like stars (i.e.,
are not ejected).  However, current star formation models (see e.g.,
Bate et al.\ 2003) show that the majority of remnant disks in brown
dwarfs have radii less than 20~AU; these calculations do not possess a
resolution sufficient to follow the fate of these disks after
ejection. Hence the possibility remains that BDs can retain a disk
even when they are ejected. With an accretion rate of
$\dot{M}=5\,10^{-12}\,M_\odot.yr^{-1}$ (Muzerolle et al.\ 2005), even
a  $5\,10^{-5}\,M_\odot$  disk would survive a few Myr, a lifetime
consistent with the age of Taurus. 


  

In this paper, we present complete photometry available on those 23 BDs in
Taurus for which {\it Spitzer} data are available, from the visible to
$70\,\mu$m (Section~\ref{sec:results}). We combine this large range of
photometry with other observations such as spectral types to further
aid in interpretation.  We sort these SEDs depending on the presence
of an infrared excess and we fit these excess-bearing sources with a
disk model (Section~\ref{sec:models}). We discuss the implications of
our results for BD formation models in Section~\ref{sec:discussion}.


\section{Observations and results \label{sec:results}}

 \subsection{Observations}
\label{subsec:obs}

Table~\ref{tab:source-data} lists  the names, spectral types,
temperatures, and the photometric measurements described in this
section for the 23 BDs for which {\em Spitzer} photometry is
available.   The observations reported here have been collected with
various instruments from the visible to the infrared range. Not
all objects have been measured in every photometric band available.

Mid-infrared photometry has been obtained with the IRAC (at
effective wavelengths of 3.6, 4.5, 5.8, and $8\,\mu$m;  Fazio et
al.\ 2004) and MIPS (24, 70, and $160\,\mu$m, Rieke et al.\
2004) instruments on board the {\it Spitzer Space Telescope}
(Werner et al.\ 2004). The {\it Spitzer} fluxes were extracted
from the mosaic images obtained as part of General Observer
program 3584 (PI: D.\ Padgett) for wide area ($\approx30\,{\rm
deg}^2$) mapping of the Taurus cloud (Guedel et al.\  2006). The
observations were carried out in February and March 2005, with
exposures at two epochs several hours apart to provide for
asteroid rejection in this low-ecliptic-latitude region.
Photometry was performed using the APPHOT and PHOTCAL packages
in IRAF.  We measured fluxes within an aperture radius of 3
pixels (3.$^{\prime\prime}$6) and applied aperture corrections
to 8 pixels (9.$^{\prime\prime}$6) for IRAC bands.  At 24
microns, we used 3 pixel (7.$^{\prime\prime}$4) apertures and
corrected the flux to 11 pixels (27.$^{\prime\prime}$0); at 70
microns, we also used 3 pixel (29.$^{\prime\prime}$5) apertures
and corrected the flux to 8 pixels (78.$^{\prime\prime}$7). The
aperture corrections from 3 pixels to 8 or 11 pixels were
derived based on the photometry of bright isolated point sources
in the mosaic images.  The fluxes were measured at each
individual epoch, then averaged to obtain the final values. We
have computed the corresponding magnitudes adopting IRAC zero
magnitude flux densities of 280.9, 179.7, 115.0, and
64.13~Jy in the 3.6, 4.5, 5.8 and 8 micron bands, respectively
(Reach et al.\ 2005); for MIPS, we used zero points of 
7.14 Jy for the 24 micron band and 0.775 Jy for the 70 micron
band, based on the MIPS Data Handbook.  
The uncertainty on the absolute calibration is 10\% for the IRAC
bands and 24~$\mu$m, and 20\% for 70~$\mu$m measurements. 


The IRAC fluxes for 23 Taurus BDs are shown in
Table~\ref{tab:source-data}. CFHT-Tau\_12 and KPNO-Tau\_9 were not
observed at all IRAC bands. Eleven sources were detected at
$24\,\mu$m, and just one source (J04442713+2512164) was detected at
$70\,\mu$m, with an uncertainty of 0.2~mag. The upper limits of
undetected sources are 0.9\,mJy at $24\,\mu$m, and $\lesssim100\,$mJy
at $70\,\mu$m, depending on the background level due to the cloud
emission. The error bars are within the size of the symbol in Figure
2, except for the 70~$\mu$m measurement.  Note that IRAC photometry
for objects KPNO-Tau 4 to 7 and GM~Tau has already been published in
Hartmann et al.\ (2005). 

We have also used the 2MASS catalog to obtain $J$, $H$ and $K_s$ band
photometry for the whole sample of Taurus BDs.  The transformation
between 2MASS photometry and absolute flux has been performed using
the zero-point fluxes from Cohen (2003), specifically 1594, 1024, and
666.7~Jy  for $J$, $H$ and $K_s$ bands, respectively.

The optical data come from several telescopes. We have obtained $I$ 
photometry using CFHT12k and Megacam cameras on the
Canada-France-Hawaii telescope for 16 BDs.  In addition, 8~BDs possess
$R$ photometry obtained with the same instruments. The main
characteristics of this photometric observations are presented in
Guieu et al.\ (2006). Additional visible photometric data have been
obtained from  Briceno et al.\ (2002), Luhman et al.\ (2003), and
Luhman (2004). 

 
{\small

\begin{table*}[htb]
\caption {Visible - infrared photometry for the 26 BDs studied in this paper. The sources with IR excess are labeled with $^*$. The data are given as magnitudes (see text for details) except for the 24 and $70\,\mu$m fluxes, given in mJy.  \label{tab:source-data}}
\begin{tabular}{ l|ll|lllll|llll|rr } \hline\hline
 Name&SpT&Av&R&I&J&H&K&[3.6]&[4.5]&[5.8]&[8.0]&24&70 \\&&&&&&&&&&&& \multicolumn{2}{c}{ mJy}
 \\ \hline
CFHT-Tau\_9$^*$&M6.25&0.91&\ldots&15.35&12.88&12.19&11.76&11.14&10.86&10.45&9.80&12.00&$<$57.20\\
KPNO-Tau\_4&M9.50&2.45&20.54&18.75&15.00&14.02&13.28&12.52&12.34&12.14&12.10&$<$0.89&$<$57.20\\
CFHT-Tau\_15&M8.25&1.30&\ldots&17.94&14.93&14.24&13.69&13.19&13.16&13.04&13.04&$<$0.89&$<$57.20\\
KPNO-Tau\_5&M7.50&0.00&19.10&15.08&12.64&11.92&11.54&11.03&10.99&10.90&10.84&$<$0.89&$<$57.20\\
KPNO-Tau\_6$^*$&M9.00&0.88&20.56&17.90&14.99&14.20&13.69&13.08&12.79&12.47&11.68&$<$1.65&$<$75.80\\
CFHT-Tau\_16&M8.50&1.51&\ldots&17.91&14.96&14.24&13.70&13.26&13.16&13.01&12.97&$<$0.89&$<$57.20\\
KPNO-Tau\_7$^*$&M8.25&0.00&\ldots&17.16&14.52&13.83&13.27&12.59&12.27&11.90&11.22&2.29&$<$57.20\\
CFHT-Tau\_13&M7.25&3.49&\ldots&17.90&14.83&13.97&13.45&12.74&12.73&12.84&12.60&$<$1.20&$<$72.20\\
CFHT-Tau\_7&M6.50&0.00&16.63&14.12&11.52&10.79&10.40&9.83&9.91&9.70&9.69&$<$1.41&$<$134.00\\
CFHT-Tau\_5&M7.50&9.22&\ldots&18.79&13.96&12.22&11.28&10.48&10.25&10.03&10.01&$<$0.94&$<$57.20\\
CFHT-Tau\_12$^*$&M6.50&3.44&\ldots&16.26&13.15&12.14&11.55&\ldots&\ldots&\ldots&\ldots&3.38&\ldots\\
CFHT-Tau\_11&M6.75&0.00&\ldots&14.88&12.53&11.94&11.59&11.13&11.09&10.96&10.96&$<$0.89&\ldots\\
KPNO-Tau\_9&M8.50&0.00&\ldots&18.76&15.48&14.66&14.19&13.51&\ldots&\ldots&\ldots&$<$0.89&\ldots\\
CFHT-Tau\_2&M7.50&0.00&\ldots&16.81&13.75&12.76&12.17&11.54&11.41&11.32&11.33&$<$0.89&$<$65.90\\
CFHT-Tau\_3&M7.75&0.00&\ldots&16.88&13.72&12.86&12.37&11.78&11.69&11.61&11.59&$<$0.89&\ldots\\
J04380083+2558572&M7.25&0.64&20.21&14.71&11.54&10.62&10.10&9.56&9.45&9.31&9.28&1.22&$<$79.00\\
J04381486+2611399$^*$&M7.25&0.00&20.33&17.84&15.18&14.13&12.98&10.77&10.19&9.66&8.91&62.80&$<$70.70\\
GM\_Tau$^*$&M6.50&4.34&\ldots&15.04&12.80&11.59&10.63&9.25&8.76&8.38&7.80&46.30&$<$103.00\\
CFHT-Tau\_6$^*$&M7.25&0.41&18.40&15.40&12.64&11.84&11.37&10.72&10.44&10.00&9.11&15.90&$<$81.50\\
CFHT-Tau\_4$^*$&M7.00&3.00&\ldots&15.78&12.17&11.01&10.33&9.48&9.06&8.58&7.79&66.00&$<$77.80\\
CFHT-Tau\_8$^*$&M6.50&1.77&19.28&16.43&13.17&12.12&11.45&10.83&10.31&9.86&9.18&16.80&$<$57.20\\
J04414825+2534304$^*$&M7.75&1.06&\ldots&17.03&13.73&12.80&12.22&11.37&10.88&10.43&9.53&18.40&$<$57.20\\
J04442713+2512164$^*$&M7.25&0.00&\ldots&\ldots&12.20&11.36&10.76&9.51&8.99&8.33&7.40&124.00&157.00\\
 \hline
\end{tabular}
\end{table*}
 }
 
All $R$ and $I$  data have been transformed to the CFHT12k camera's
Cousins system (see Guieu et al.\ 2006;  Briceno et al.\ 2002) before
conversion to absolute fluxes. The magnitudes are listed in
Table~\ref{tab:source-data} while the absolute fluxes are used in SED
fitting (see Section~\ref{sec:analysis}). 

\subsection{SEDs and infrared excesses}

\begin{figure*}[htb]
\includegraphics{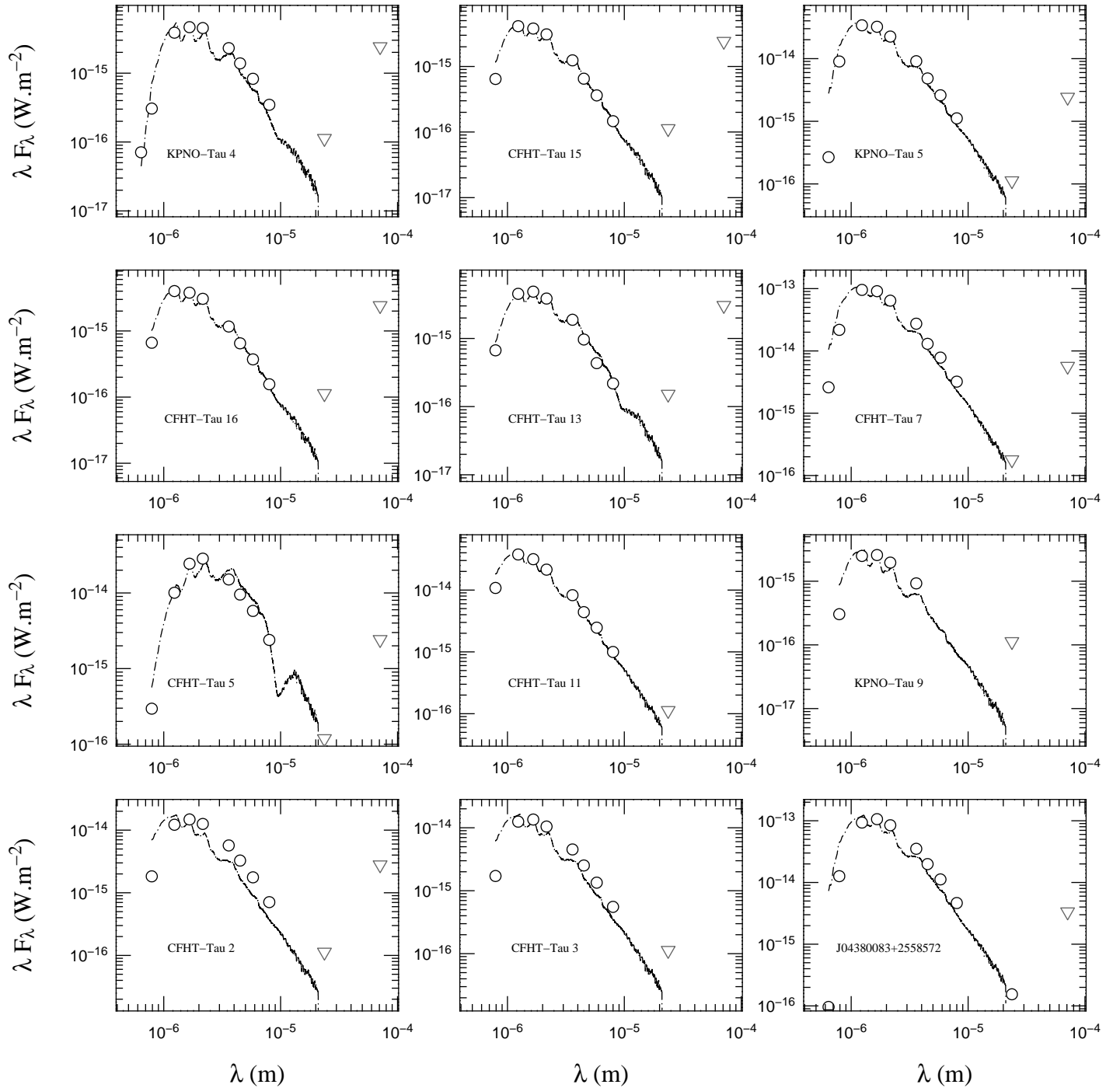}
\caption{ Mosaic of SEDs for BDs classified as having no excess
($\lambda F_\lambda$ in W m$^{-2}$). Empty triangles denote {\em
Spitzer} upper limits. Dot-dashed line: DUSTY models from Allard et
al.\ (2000) fitted on the visible-NIR range.}
\label{fig:sed1}
\end{figure*}

\begin{figure*}[htb]
\includegraphics{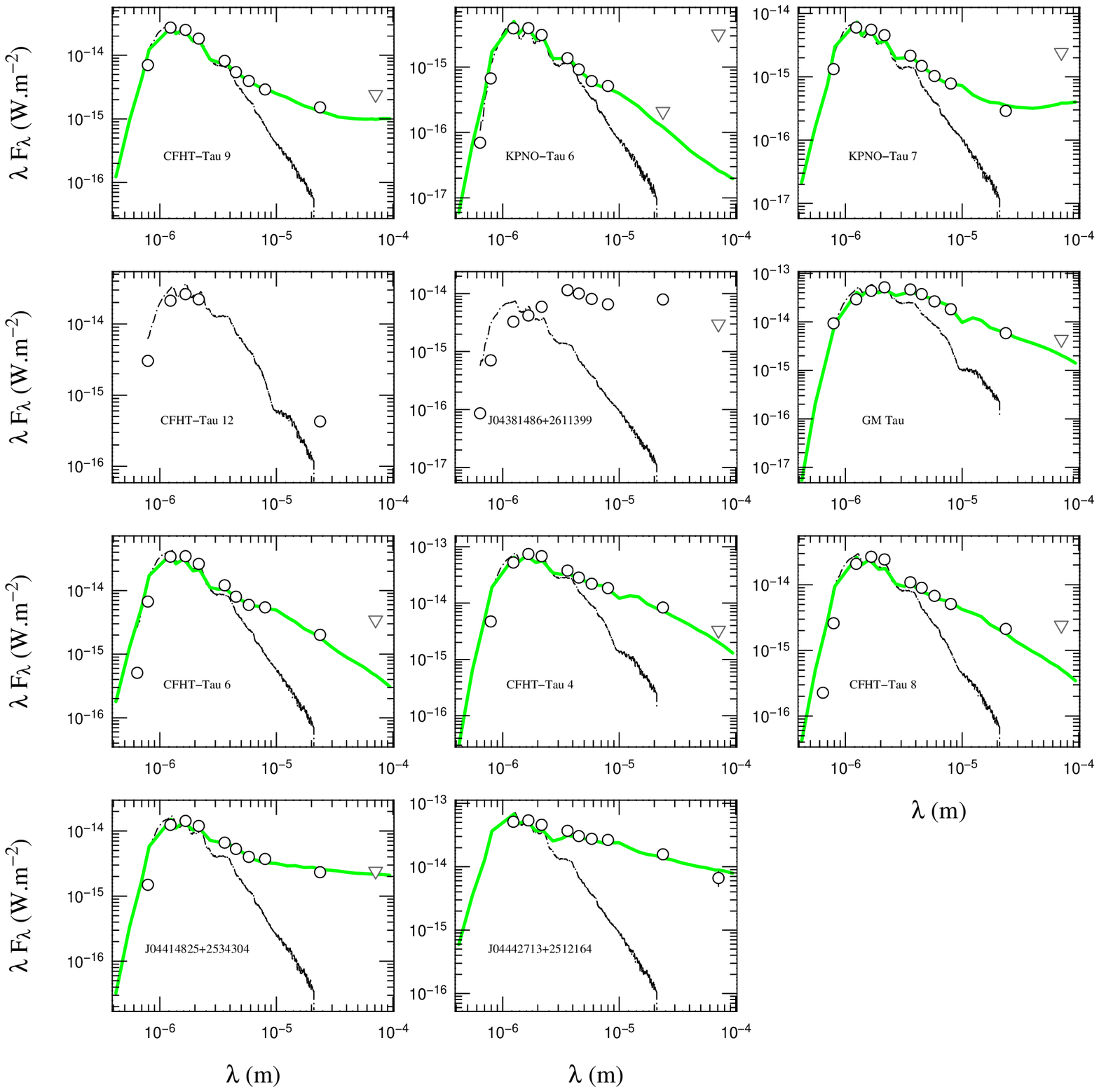}
\caption{Same as figure \ref{fig:sed1} for BDs with infrared excesses.
The green line traces our SED models; see text and
Section~\ref{sec:models} for details.} 
\label{fig:sed2}
\end{figure*}

We have plotted the SEDs in Figures \ref{fig:sed1} and \ref{fig:sed2}
for all Taurus BDs with available IRAC photometry. The SEDs are
sorted depending on the presence of an IR excess determined as
explained below. On the same plots, we have superimposed a
photospheric fit using the DUSTY models from Allard et al.\ (2000). 
Each model has been fit using the BD's temperature and visual
absorption previously published in the literature combined with the
Draine (2003a, 2003b, 2003c) extinction law.  Following Natta and
Testi (2001), who show that BD disk excess emission becomes clearly
detectable only longward of $3\,\mu$m, we have fit the DUSTY models to
the available data points just in the $R$, $I_c$, $J$, $H$ and $K_s$
bands. 
We decide that a given source has an IR excess when the IRAC
photometry exceeds the photospheric SED by more than one sigma.
For some sources (e.g., CFHT-Tau~2), the IRAC measurements fall 
somewhat above the SED but the slope of the points is the same
as the underlying photosphere. In such cases, we conclude that
it is more likely that the shift comes from a small fit mismatch
rather than from a real IR excess. The only exception to this
rule is CFHT-Tau~12, which was not observed with IRAC, but shows
a confirmed detection at $24\,\mu$m. We have thus placed this BD
in the IR excess list.

As a preliminary conclusion, we find that 11 objects out of 23
have a significant IR excess, suggesting that $48\%\pm 14\%$ of
Taurus BDs possess disks. We computed the uncertainty on this
percentage using simple Poisson statistics on the number of BDs
with disks. This proportion is very similar to the one of
classical T~Tauris (CTTs) among T~Tauri stars in Taurus
(Hartmann et al.\ 2005), and is fully consistent with the
proportion of BDs with disks found by Luhman et al.\ (2005) in
IC\,348 and Cha~I.  In section~\ref{sec:models}, we will present
the models used to fit the SEDs with infrared excesses.  These
fits are superimposed on the objects' SEDs plotted in Figure~2.
We could only fit 9 out of 11 SEDs for reasons explained in
Section~\ref{sec:models}.  In the following paragraphs, we
explore other physical parameters to study the presence of disks
around our objects. 

 \subsection{Spitzer color-color diagrams}
 \label{sub:spi-colcol}
 
\begin{figure}[htb]
\includegraphics[width=\hsize]{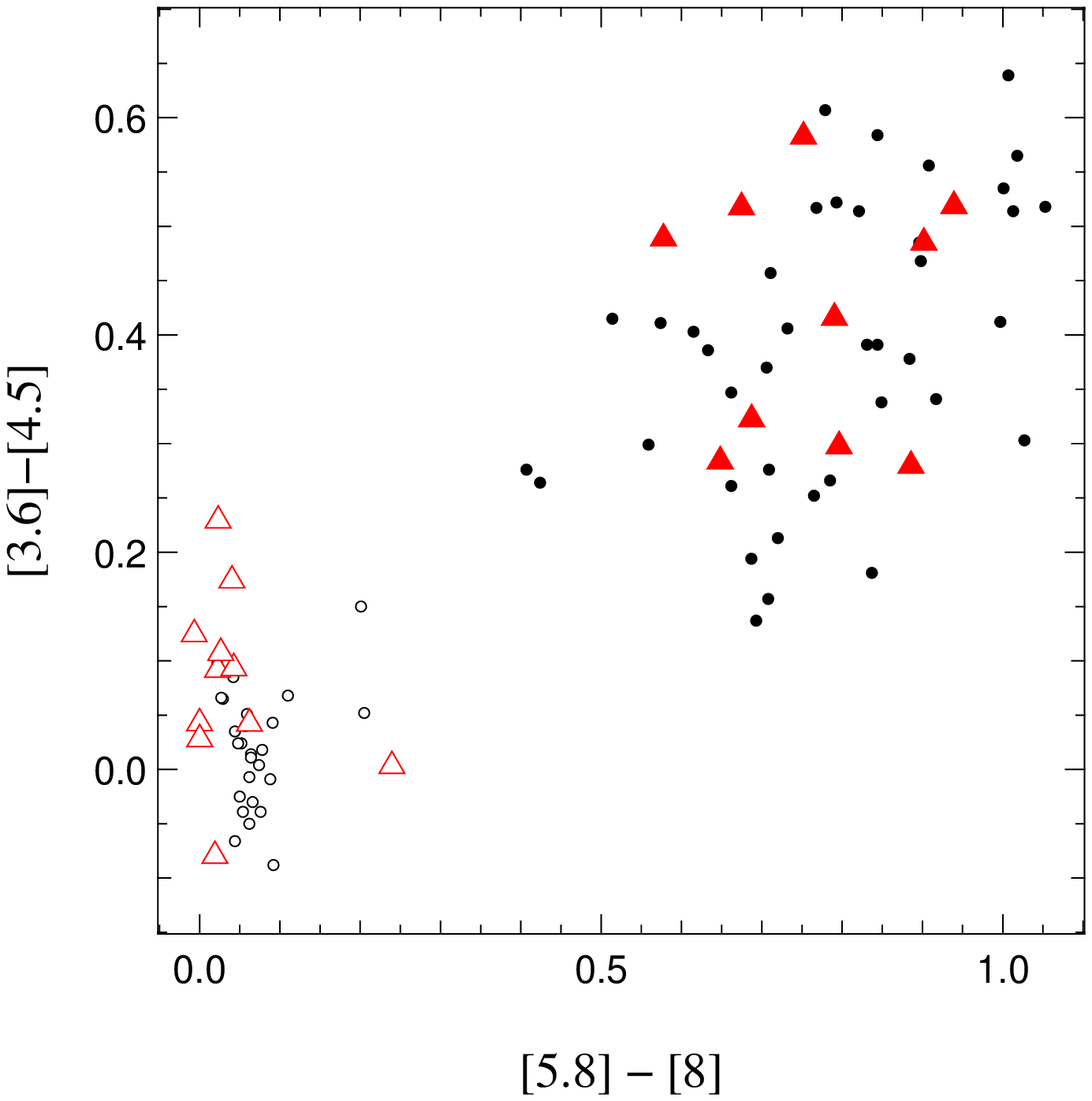}
\caption{[3.6]$-$[4.5] vs.\ [5.8]$-$[8]  color-color diagram of the
BDs studied in this work, superimposed on the TTS from  Hartman et
al.\ (2005). Filled / empty triangles: BDs with / without IR excesses;
filled circles: class~II (CTTS); and empty circles: class~III (WTTS).
The typical uncertainty on Spitzer color indices is 0.05 (see
sec.~\ref{subsec:obs}) \label{fig:i1234}}
\end{figure}

In the previous section, the presence or absence of a disk
around a BD is inferred more from the slope of the IRAC data
than from the actual value of the IRAC data compared to the
photospheric fit. To confirm our results, we have plotted in
Figure~\ref{fig:i1234} the {\it Spitzer}  [3.6]$-$[4.5] {\it
vs.} [5.8]$-$[8] color-color indices for all the BDs where the
photometry is available. The data are plotted as full / empty
triangles depending on the presence / absence of an IR excess in the SED. 
We find that the Taurus BDs are essentially plotted in two distinct
regions, identical to the ones found for T~Tauri stars (TTS)
by Hartmann et al.\ (2005).  
We interpret the distinction between these two regions as due to the
presence or absence of circumstellar dust, in a way consistent with the 
classification adopted to sort the SEDs in
Figures~\ref{fig:sed1} and \ref{fig:sed2}. On the same figure,
we have superimposed the data points from Hartmann et al.\
(2005)  as open / filled circles for Taurus class II/III stars.
There appears to be no clear segregation between BDs and TTS
colors when they have a disk. In contrast, in the lower left
corner of the plot, BDs without disks tend to be redder on the
average than their WTTS equivalents as measured by the
[3.6]$-$[4.5] index.  
 Figure~\ref{fig:i1234} shows that CTTs and BDs with infrared
excesses appear indistinguishable, consistent with their
emission being dominated by their disks in both cases, while the
WTTs and BDs without disks can be distinguished by their
photospheric colors. 
 
\begin{figure}[htb]
\includegraphics[width=\hsize]{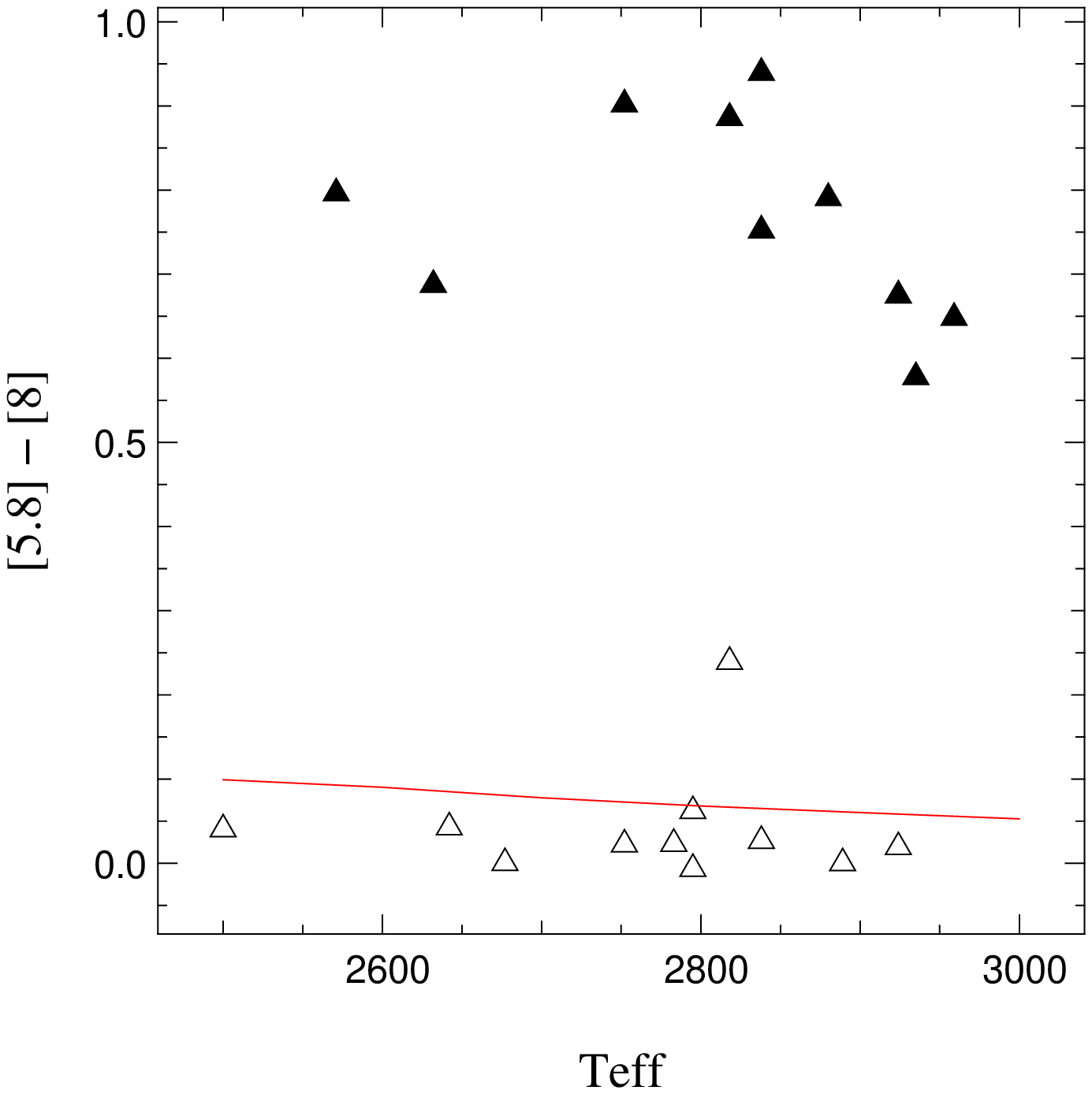}
\caption{[5.8]$-$[8] $\mu$m color index of the disk emission versus
effective temperature of the corresponding central object. The solid
line delineates the  model computed using the Allard et al.\ (2000)
photosphere. Open / filled triangles denote BD without / with IR
excess. The typical uncertainty on the Spitzer color indexes is
0.05.} 
\label{fig:excess}
\end{figure}

%
%
In Figure \ref{fig:excess} we have plotted the IRAC [5.8]$-$[8]
color index versus the underlying  effective temperature of the
corresponding photosphere. Again, there is a clear gap of about
0.7 mag between the BDs with and without disks, with no clear
dependence of the color index on the central object effective
temperature, hence its mass, at a given age.


\subsection{Accretion signatures}
 
\begin{figure}
\includegraphics[width=\hsize]{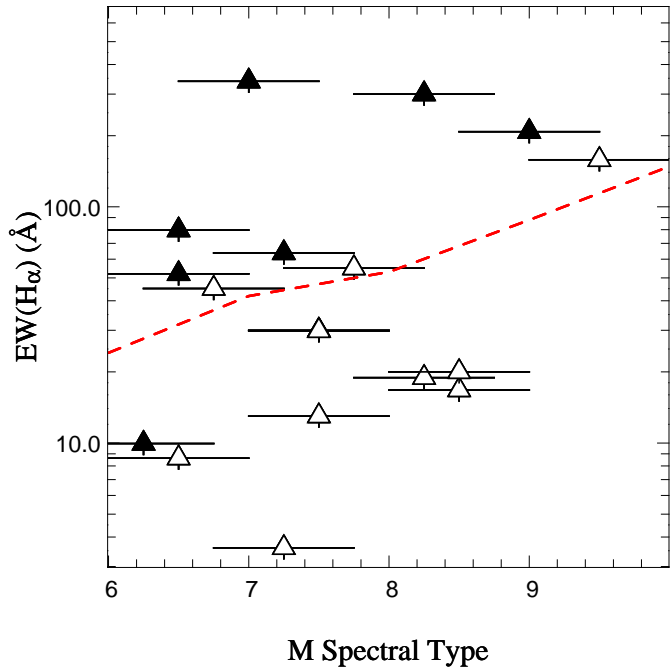}
\caption {H$\alpha$ equivalent width versus M 6-9 spectral type for
all the BDs in our sample where data are available. Full triangle: BD
with a disk; empty triangle: BD without a disk.}
\label{fig:halpha}
\end{figure}

In Figure~\ref{fig:halpha}, we have plotted the H$\alpha$
equivalent width (EW; Guieu et al.\ 2006; Brice\~no et al.\ 2002)
from spectra obtained with moderate resolution ($R\approx 1000$)
versus spectral type for all BDs where {\it Spitzer} photometry
and EW(H$\alpha$) are available (filled triangles: BDs with
disks; empty triangles: BDs without disks).  The empirical
CTTS/WTTS boundary extended to substellar analogs, defined by
Barrado y Navascu\'es and Mart\'in (2003), is delineated by  the
dashed line. This boundary is defined by the saturation limit
for chromospheric activity ($L_{{\rm H}\alpha}/L_{bol} = -3.3$).

There is a general agreement between the accretion / non accretion
limit from Barrado y Navascu\'es and Mart\'in (2003) and our disk / no
disk criterion. All but one of the BDs with disks have an H$\alpha$
emission level in excess of that expected from chromospheric activity,
suggesting that most BDs with disks are experiencing an accretion
phase, or a jet.  Moreover, although we are dealing with small
numbers, there appears to be two groups of accreting BDs in our data: 
BDs with EW$({\rm H}\alpha)>300\,$\AA\ are strong accretors, while
objects with  EW$({\rm H}\alpha)<100\,$\AA\  may have stopped
significant accretion and be surrounded by more passive disks, analogs
to the one described in McCabe et al.\ (2006). 

As the H$\alpha$ EW might not provide an unambiguous accretion
status of a given object, we have compared our results with
those from Mohanty et al.\ (2005). They study the accretion in
BDs using the H$\alpha$ 10\% width.  Among the 23 sources
studied in this paper, 9 were observed by Mohanty et al.\
(2005).  We find that among our 11 sources classified as BDs
with  disks, 3 are found to be accreting by Mohanty et al., and
one is found to be a {\em possible} accretor with an H$\alpha$ 
10\% width = 150 km s$^{-1}$  (their accretion limit is at
200~km s$^{-1}$). Moreover, this latter object is CFHT-Tau~4,
which was found by Pascucci et al.\ (2003) to harbor a disk, as
shown by its mm emission. In our H$\alpha$ measurements, we find
that CFHT-Tau~4 has an H$\alpha$ equivalent width of 300\,km
s$^{-1}$.  Additionally, among the 12 sources we classify as BDs
without disks, 4 are found to be non-accretors by Mohanty et
al.\ (2005), and one is a possible accretor (KPNO-Tau~4). Using
this limited overlap between the Mohanty et al.\ (2005) data and
ours, we can conclude that there is a correlation between our
accretion classification and theirs. If we ignore the objects
that appear as intermediate, or passive, the presence of a disk
inferred from the IR excess is thus highly correlated to the
presence of accretion. 

As a word of speculation, it could be noted that there is a small
mismatch between the accretion limit and our three low EW(H$\alpha$)
objects; the difference falls within the error bars, although the
shift is in the same direction for all three objects. Rather
than speculating about possible remnant accretion in BDs without strong disks, it could be
possible that the extension in the BD domain of the boundary drawn by
Barrado y Navascu\'es and Mart\'in (2003) has to be offset by
$+20-30\,\%$ to fit our new data points.  Indeed, if we raise their
limit by this amount, it nicely follows our BDs without  disks (open
triangles) leaving all the low-accreting BDs with disks below the
chromospheric activity limit. 


%

\subsection{Brown dwarfs spatial distribution}
\label{subsec:bd-spat}

\begin{figure}
\includegraphics[height=\hsize]{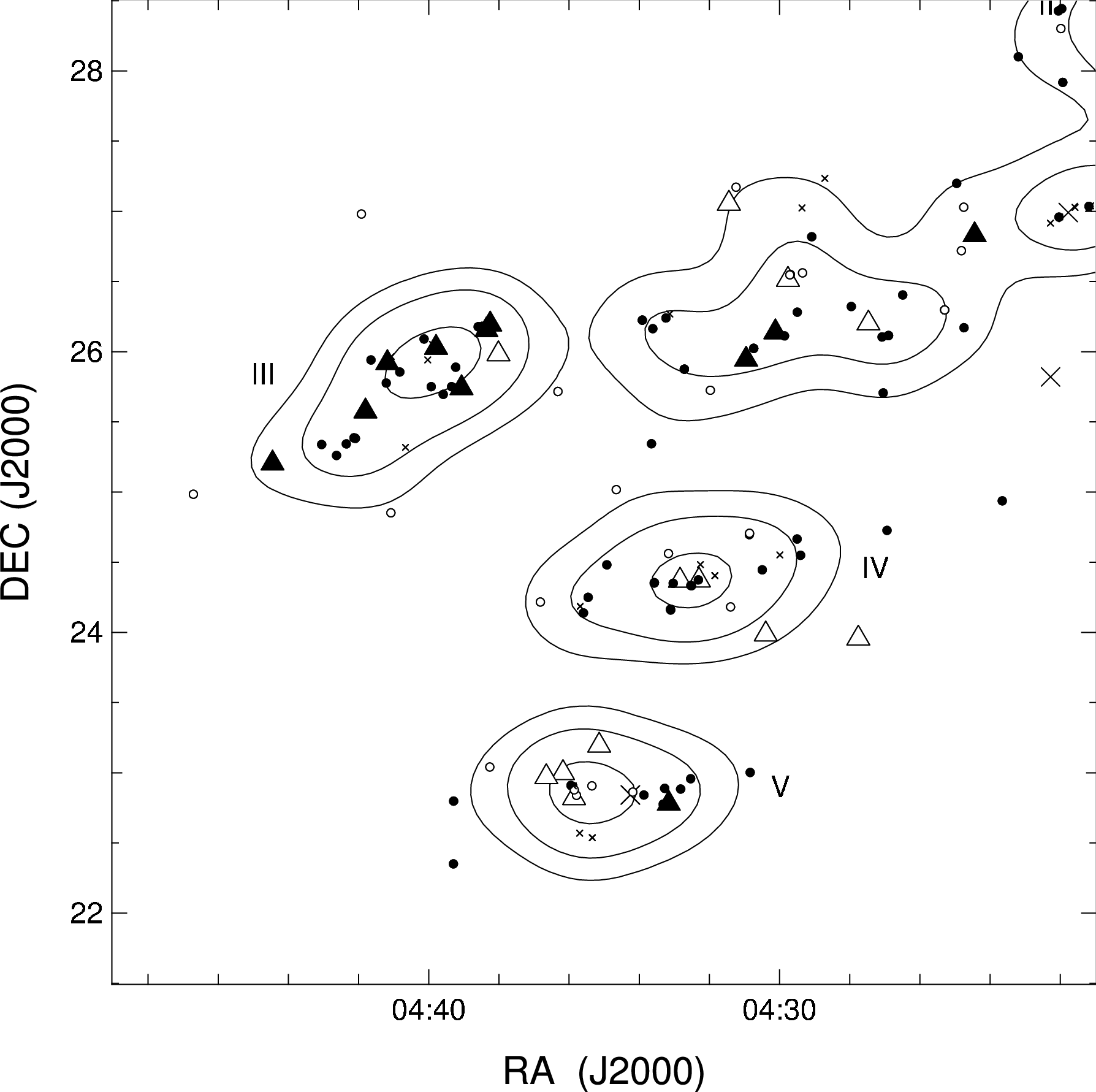}
\caption {Spatial distribution of Taurus BDs. Filled triangles
are BDs with disks; open triangles are BDs without disks. Filled
and open circles denote class I+II and III  TTauri stars (TTS).
The few crosses mark stars that could not be classified.  The
solid lines delineate the 3, 6, and 12 star/deg$^2$ isodensity
contours. }
\label{fig:spatial-dist}
\end{figure}

Figure~\ref{fig:spatial-dist} shows the spatial distribution of
our objects, superimposed on the aggregates defined by Gomez et
al.\ (1993) plotted as stellar isodensity peaks; we label them
from III to V accordingly (the aggregate in the upper right part
of the figure was not labeled by Gomez et al.\ 1993). We plot
isodensity contours at 3, 6 and 12 stars per square degree.  BDs
with disks are plotted as filled triangles and BDs without disks
are plotted as open triangles. We also show the positions of the
class I+II and III (C/W) T~Tauri stars of the Taurus population,
as compiled from Kenyon and Hartmann (1995), White and Ghez
(2001), Hartmann et al.\ (2005), and Andrews and William (2005).
We find that, although globally, the fraction of BDs with disks
is about 50\%, BDs with and without disks are not distributed regularly. 
For instance, aggregate III has 7 out of 8 BDs with disks, while
aggregate V has 4 out of 5 BDs without disks, the remaining one being
BD CFHT-Tau~12, which has the lowest IR excess from all the sample at
$24\,\mu$m.
We have also counted the proportions of class I+II and III
(C/W)TTs in the aggregates, inside the 3 star/deg$^{2}$
contour. The proportions of CTTs and BDs with disks in the
various aggregates where this information is available are
listed in Table~\ref{tab:agg-data}.  We have also listed the
(binomial) probability to get the aggregate BD disk
proportion if a given BD has a 48\% probability to get
a disk. 
Given the uncertainty on the proportion of BDs with disks, the
numbers listed in Table~\ref{tab:agg-data} are not absolutely
inconsistent with an overall 50\% disk frequency, although with
rather low probability. Indeed, if the probability for a BD to
have a disk falls down to its lowest value (34\%), there is a
20\% chance to find the repartition (0/4) in aggregate IV. 
However, this ensemble of results could be evidence of
different physical conditions in the various aggregates. We come
back to this point in Section~\ref{sec:discussion}. 

In order to check if there is a difference of spatial position
of the BDs with and without disks relative to the stellar
aggregates, we have checked a series of estimators:  {\em i}) We
have computed the average distance to the nearest star for BDs
with and without disks, and found no significant difference
between the two. {\em ii}) The average distance from a BD with
and without disk to its nearest aggregate center are similar. 
{\em iii}) We have also computed the weighted stellar density
where each BD stands, using the Kernel method (Silverman 1986),
and we find no significant difference between BDs with and
without disks. 

%

\begin{table}[htb]
\caption {CTTS and BDs with disk proportions in aggregates where
the information is available. The number listed in the third
line are the binomial probabilities to get the observed BD
repartition, computed with $p_{\rm BD}({\rm disk}) =
0.48$.\label{tab:agg-data}}
\begin{tabular}{llllll} \hline\hline
  Aggregate  & III & IV & V \\
 CTTS   &   15/18 & 15/19& 8/13  \\
 BD disks   &  7/8 & 0/4  &  0-1/5\\
 Probability & 2.4\% & 7\% & 3.8-17 \%
 \\ \hline
  \hline
\end{tabular}
\end{table}

\section{Disk models}
\label{sec:models}

In order to estimate the distribution and the amount of
material present in the disks around the BDs showing an IR
excess, we have used a 3D Monte-Carlo continuum radiative
transfer code (MCFOST, see Pinte et al.\ 2006), to model the
SEDs of the BDs with IR excess.  Our model includes multiple
scattering with 
passive dust heating, assuming radiative equilibrium
and continuum thermal re-emission.

\subsection{Dust distribution in the disk}

We use a density distribution with a Gaussian vertical profile
$\rho(r,z)=\rho_0(r)\,\exp(-z^2/2\,h^2(r))$, assuming a
vertically isothermal, hydrostatic, non self-gravitating disk.
We use power-law distributions for the surface density
$\Sigma(r) = \Sigma_0\,(r/r_0)^{\alpha}$  and the scale height $
h(r) = h_0\, (r/r_0)^{\beta}$ where $r$ is the radial coordinate
in the equatorial plane, $h_0$ the scale height at the radius
$r_0 = r_{in}$. The disk extends from an inner radius  $r_{\rm
in}$  to an outer limit radius $r_{\rm out}$. The central star
is represented by a sphere radiating uniformly with  photosphere
parameters extracted from the literature (Guieu et al.\ 2006
and references therein) and the corresponding synthetic brown
dwarf spectra of Allard et al.\ (2000) shown in
Figure~\ref{fig:sed1} and \ref{fig:sed2}. 

\subsection{Dust properties}

We consider homogeneous spherical grains and we use the
dielectric constants described by Mathis \& Whiffen (1989) in
their model A, with typical interstellar medium values. The
differential grain size distribution is given by $\mathrm{d}n(a)
\propto a^{-3.7}\,\mathrm{d}a $ with grain sizes between
$a_{\mathrm{min}} = 0.03 \mu$m and  $a_{\mathrm{max}} = 1 \mu$m.
The mean grain density is $0.5$ g cm$^{-3}$ to account for
fluffiness. Extinction and scattering opacities, scattering
phase functions and Mueller matrices are calculated using Mie
theory. Dust and gas are assumed to be perfectly mixed and grain
properties are taken to be independent of position within the
disk. The total disk mass (gas+dust) is fixed at $1\,M_J$,
assuming a gas to dust mass ratio of 100. 

\subsection{Model fitting}
\label{sec:analysis}

The fits were performed using a grid of SED models built by
variation of 5 free parameters whose values are listed in Table~\ref{tab:parameter_space}.

\begin{figure}
\includegraphics[width=\hsize]{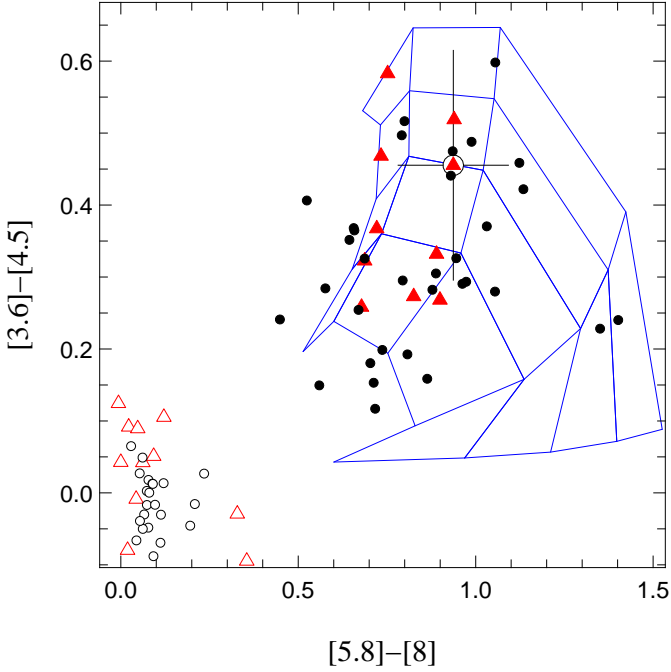}
\caption {Dereddened BD + TTS  superimposed on disk models in a
color-color diagram, with $h_o$ and $h_o/r_{\rm in}$ taking the
values listed in Table~\ref{tab:parameter_space}. Starting from
the lower left point of the grid $(0.5,0.2)$, $h_o/r_{\rm in}$
varies from $0.02$ to $0.1$ upward, while $r_{\rm in}$ varies
from $0.015$ to $0.3\,$AU to the right.}
\label{fig:pts-models}
\end{figure}

\begin{table}[htb]
  \caption{Parameter range of the models computed in this
paper.\label{tab:parameter_space}}
  \begin{tabular}{llllll}
    Parameter & \multicolumn{5}{c}{range values}\\
    \hline
    $r_\mathrm{in} ({\rm AU})$   & 0.015   & 0.032  & 0.067  & 0.14  & 0.3  \\
    $\beta$                              & 0.0       & 1.0     & 1.125   & 1.25 &        \\
    $\alpha$                             & -0.5      & -1       & -1.5     &        &         \\
    $h_0/r_{in}$                        & 0.02     & 0.04    &  0.06    & 0.08 &  0.1   \\ 
    $cos(i)$  &  \multicolumn{5}{l}{from 0.05 to 0.95 by steps of 0.1}   \\ 
    \hline
  \end{tabular}
\end{table}
For each object, the models are sorted following a
pseudo-$\chi^2$ minimisation. We decide that a parameter value
is acceptable when the corresponding $\chi^2$ value is less than
twice the $\chi^2$ of the best model. For each model SED, we
compute its {\em Spitzer} colors in the 3.6, 4.5, 5.8 and 8
$\mu$m bands and we plot them in the [3.6]$-$[4.5] vs.
[5.8]$-$[8] color color diagram.  Using this diagram as a
diagnostic tool, we find that when we vary $\beta$, $\alpha$,
and $\cos (i)$, the corresponding color points are spread
randomly across the diagram. In contrast, the variation of
$r_{\rm in}$ and $h_o/r_{\rm in}$ define a coherent grid across
the diagram.  Figure~\ref{fig:pts-models} shows the deredened
TTS and BD color points superimposed on such a model grid
obtained for $h_o/r_{\rm in}$ varying in the range listed in
Table~\ref{tab:parameter_space} ($5\times 5$ values). Each point
is computed for a given $(h_o,r_{\rm in})$ couple, with all the
results due to other parameters variations averaged.  Of course,
the result of a given model depends on the central object used
to compute the surrounding disk parameters. In 
Figure~\ref{fig:pts-models}, we show all the objects'
color-color points but only one model grid, computed using the
central object with corresponding color-color point encircled. 
 
In regard to the flaring exponent, we find that none of our
sources can be fitted convincingly by a flat disk model
($\beta=0$) or by a highly flaring model ($\beta=1.25$).  In
Table \ref{tab:param-list} we list the values obtained by our
fitting procedure described above.

\begin{table}[htb]
\caption {Best fit results for all the BDs with IR excesses; the
sources are listed in the same order as they are presented in
Figure~\ref{fig:sed2}\label{tab:param-list}}.
\begin{tabular}{lccc} 
  Object              & $h_0/r_{in}$ & $r_{in}$ (AU)   & $\beta$   \\ \hline
CFHT-Tau 9        & 0.02           & 0.032-0.067 & 1.125	   \\
KPNO-Tau 6        & 0.02           &  0.032      & 1.0-1.125 \\ 
KPNO-Tau 7        & 0.04          & 0.015       & 1.0-1.125  \\
GM Tau              & 0.06-0.1     & 0.015-0.067 & 1.0	   \\
CFHT-Tau 6        & 0.04           & 0.067-0.14  & 1.0-1.125 \\
CFHT-Tau 4        & 0.04-0.06    & 0.14-0.3    & 1.0	   \\
CFHT-Tau 8        & 0.04-0.06    & 0.015-0.067 & 1.0-1.125 \\
J04414825+2534304 & 0.06       & 0.015-0.032 & 1.125	   \\
J04442713+2512164  &  0.1      & 0.032-0.067 & 1.125 \\ \hline
\end{tabular}
\end{table}

Our disk fitting procedure does not address the entire disk
parameter space.  Indeed, even the $24\,\mu$m {\em Spitzer}
photometry only probes the inner part of the disk on a scale 
$\sim1$~AU, so that most of the disk mass remains hidden in the
outer parts of the disk.  In our disk model, 75\% of the disk
energy comes from the inner 1~AU and 90\% comes from the inner
3~AU.  We stress that our model is able to compute the SED up to
$\lambda\approx 1\,$mm, where the disk is optically thin, so
future mm measurements of the disks around our BD sample will be
very valuable to estimate their mass.  However, our results show
that for all the sources that we have fitted,  we can rule out a
disk structure without flaring, showing that in all the BD disks
sampled in this study, the disk retains a significant amount of
gas. This is consistent with the level of accretion observed in
almost all  BDs with infrared excess. 

In our modeling, the disk structure is described by parametric
laws. Nevertheless, the dust scale height inferred from the best
models was compared with the hydrostatic scale height (computed
from the disk temperature and corresponding to the gas scale 
height). The dust scale height is marginally smaller than, but
compatible with, the hydrostatic scale height.

The fitted SEDs are superimposed on the objects' SEDs in Figure~2.
Only 9 fits are displayed there because we could not find a
satisfactory fit result either for CFHT-Tau~12 or for
J0438+2611. For the former object, this is because of the lack
of {\em Spitzer} photometry. For the latter, the SED of this
object is the only one in the sample to show such a rise toward
longer wavelengths. We interpret this behavior as being due to
the peculiar orientation of the disk around the BD, close to
edge-on (Luhman 2004). Moreover, if it were not for the higher
emission in the $10-100\,\mu$m domain, this source is the
weakest one of the sample, consistent with a thick disk
occulting the central object. 

%
\subsection{Comparison with other models}

Strictly speaking, our study concerns the proportion of BDs with
{\em inner disks} only. Here, we compare our results with the
ones of  Scholz et al.\ (2006) who have surveyed  the 1.3\,mm
emission of Taurus BDs to study the properties of their colder
(hence farther) disks. Out of 12 objects in common between their
study and ours, only two show NIR IRAC excess emission without
outer cold mm emission.  All the BD with cold outer mm emission 
have IRAC emission in excess over the photosphere model, hence
were classified as "BD with disk" in our study.   We have
compared our disk parameters with the ones derived by Scholz et
al.\ (2006) from their model of the objects CFHT-Tau~4,
CFHT-Tau~6, and J044427+2512.  When Scholz et al.\ (2006) detect
a disk, they measure a disk mass between 0.4 and 1.2~$M_J$,
i.e., a range of values consistent with our choice to fix the
disk mass at $1\,M_J$ (see section~\ref{sec:models}). Similarly,
they use an index $\beta = 1.15$ when we find values ranging
from $1.0$ to $1.125$. They use an external radius $R_D=300\,$AU
when we use $R_D=100\,$AU, again a consistent value. If we
compute the disk scale height at the stellar radius, the
relative difference with theirs is less than 25\%.  Finally, if
we add the Scholz et al.\ mm point to our SED when it is
available, we find that our model SED matches  this new
measurement to within a factor of 2. Given the remaining 
uncertainties, we estimate that our models are consistent with
Scholz et al.\ results.

\section{discussion}
\subsection{BD distribution relative to stars}

In section~\ref{subsec:bd-spat}, we found that there is no
significant difference between BDs with and without disks
relative to the stars.  If BDs without disks are actually
ejected objects, our data suggest that they have not had the
time to travel away from their parent aggregate, or the
aggregate stars have also been ejected with a similar velocity
dispersion, as described in  Bate \& Bonnel (2005).  It is
also possible that for low density agregates such as the one
found in Taurus, the comparison of the BD spatial distribution
relative to the surrounding stellar population is not always a
discriminant diagnostic.  Goodwin et al.\ (2005) have modeled
the evolution of low density cores and find that in some cases,
the differences between the two distributions (stars and BDs)
can disappear.

\subsection{Spatial distribution of BD with and without disk}
\label{sec:discussion}
 



We find that the global proportion of BDs with and without 
disks in Taurus is similar to the C/W TTS global proportion in
the same region.  If disks are a robust tracer of the main star
formation route, this shows that the BD formation process has a
lot in common with that of the stars, and thus ejection by
itself cannot be invoked to explain a possible difference
between BD and stellar formation. The fact that this global
proportion is also very similar to the one found by Luhman et
al.\ (2005) in IC348 and Cha~I star forming regions with
different stellar density and physical parameters but similar
ages, shows that the resulting proportion of BD with disks in
a given region does not depend directly on the local stellar
density, hence on the local physical parameters, or on the
outcome of possible ejections.   

One of the most intriguing result of our study is the strong
variation of the proportion of BD with disks among the Taurus
aggregates, with a global proportion very close to 50\% at large
scale. If on average there is a 50\% probability for a BD to
harbor a disk, then the fact that the aggregate number III
harbors 7 out of 8 BDs with disks is quite improbable (3\%).  On
the other hand, in aggregate~V, 4 out of 5 BDs are found without
disks. Moreover the remaining BD in this latter aggregate  is
CFHT-Tau\_12, which possesses the lowest IR excess of all the
sample with disks. 


In order to check if this difference could be an age effect, we
have used an HR diagram to compute the age of the objects
present in all the aggregates. The ages are spread between 1 and
10 Myr, but on the average, the objects in aggregate III do not
appear significantly younger than the ones in aggregate IV.
Thus  the higher proportion of BDs with disks in aggregate~III
can not be explained by an time evolution effect alone.

On the other hand, aggregates III and IV have similar stellar
densities and have opposite BDs repartitions with respect to the
presence of a disk.  This is another clue that similar stellar
densities can result in different outcomes with respect to the
proportion of BDs with disks.  Either there can be large
fluctations in the ejection rate with such low stellar
densities, or the ejection process itself can  have  large 
efficiency variations in removing BD  disks.

%

Goodwin et al. (2005) have modeled the evolution of cores
containing small numbers of stars and BDs, as they
eject their lowest-mass members. Two of their findings appear to
be relevant here: {\em i)}  the spatial distribution differences
between stars and BDs can disappear depending on the
initial clustering conditions; {\em ii)} for clusters of young
age such as Taurus, a significant difference in the spatial
distributions of stars and BDs is seen in only 1 out of
5 simulations. Then we can speculate that numerical simulations
of low density aggregates with a few tens of stars can produce
situations where BDs may lose their disks in short times, but
also can retain their disk for long times. In order to be compared to
our results, such simulations should be able to follow the fate
of very small disks. 

 \begin{figure}
\includegraphics[width=\hsize]{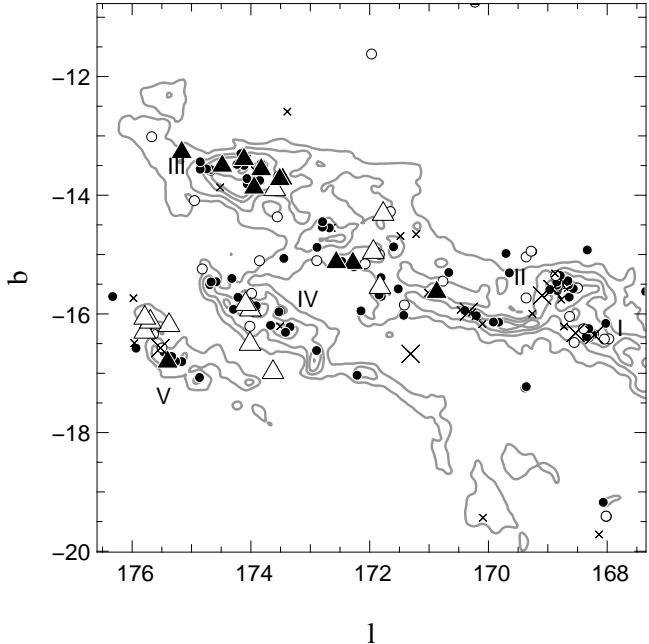}
\caption {BDs (with known {\em Spitzer} fluxes) and TTS 
superimposed on $^{13}$CO emission, plotted in galactic
coordinates. Contours are 2, 4, 6 \& 8 K km s$^{-1}$ integrated
area.}
\label{fig:fil-13co}
\end{figure}    

%
%

\subsection{Brown dwarfs and the molecular gas in Taurus}

In order to study a possibly more discriminant diagnostic than
the  nearest neighbour distance for stars and BDs, we have
considered the spatial distribution of BDs (with and without
disks) relative to the underlying cluster gas.  In
Figure~\ref{fig:fil-13co}, we have plotted the $^{13}$CO map
adapted from Mizuno et al.\ (1995) superimposed on the BD and
C/W-TTS populations in galactic coordinates. Three main
filaments appear on this figure, showing that the odd BD
repartition might in fact be related to their position relative
to the filaments.  Lepine \& Duvert (1994) have proposed that
the Taurus filaments result from a collision between a high
velocity cloud and the galactic plane. If the corresponding wave
proceeds from $b=0$,  then the leading southern filaments are
somewhat older than the northern denser, trailing one where
stellar formation would  be more recent, consistent with a
higher proportion of CTTs and BDs with infrared excess in this
filament. Note that  all the BDs found to have a  mm disk by
Scholtz et al.\ (2006)  are also found in the northern filament,
with a concentration of such BDs in aggregate~III.  However, the
projected distance between the filaments corresponds to less
than $10^6\,$yr for a wave traveling  at a few km/s, a time
difference possibly too short for significant evolution effects
to take place. This evolution problem is even more crucial if
one considers that the unnamed aggregate between II and III has
a BD disk frequency of 50\%.  Once again, a time evolution
effect by itself cannot be invoked to explain the repartition of
BDs with and without disks in the Taurus filaments.  We have
compared the level of $^{13}$CO emission at the position of the
BDs with and without  disks and we find that BDs with disks are
placed at positions where the average $^{13}$CO emission ($4.8 
\pm 2.8\,{\rm K km s}^{-1}$) is almost two times larger  than for
the BDs without  disks ($2.5\pm 2.4\,{\rm K km s}^{-1}$), although
with large uncertainties. This result mainly comes from the fact
that the BDs with disks are almost all found in the northern
filament where the $^{13}$CO emission is the strongest. As
$^{13}$CO emission is optically thin, this could suggest that
aggregates where the gas is denser have produced objects with
larger disks. 

Finally, if the three filaments yet correspond to a time
evolution series from the supposedly youngest III  to the oldest
V, there is a possibility that we are witnessing  a time
evolution difference between BDs with disks and T~Tauri stars
with disks. The C/W-TTS proportion is high in filaments III and
IV (consistent with  them being younger) and similar to the one
over the Taurus cloud ($\approx 50\%$) in filament V.  However,
the BDs in filament IV have already lost all their disks when
TTS retain a significant amount of disks there. If this effect
is real, this could be an evidence of a central object mass
effect on its disk lifetime. A similar effect has been reported
by Lada et al.\ (2006) in IC 348 where the disk fraction appears
to be a function of spectral type and stellar mass. Although
appealing,  strong issues remain:  {\em i)} we could not clearly
find a significant age difference between the objects found
in the three filament;  {\em ii)} this result is in
contradiction with a disk lifetime $\propto 1/M_*$ (e.g.,
Alexander \& Armitage 2006). Clearly more observational and
theoretical work is needed to conclude about the question of a
different lifetime in BD disks relative to TTS disks.

\section{Conclusions}

Using 0.6-70\,$\mu$m photometry, we have studied  the disk
properties  around 23 young BDs in Taurus. Using optical to
2\,$\mu$m data, we fit a photosphere model from Allard et al.\
(2000)  to all our objects. From their SEDs, we distinguish BDs 
with IR excesses (strongly suspected to have a disk) from BDs with
no excess (BDs without disks). For the BDs showing an IR excess
longward of 3\,$\mu$m, we have fit a disk model (Pinte et al.\
2006) and derived the main disk parameters. 

We find that $11/23=48\%\pm14\%$ of Taurus BDs show a
circumstellar disk signature.  This ratio is similar to the one
observed among  CTTS/WTTS in Taurus (Hartmann et al.\ 2005), and
to recent results on BDs from Luhman et al.\ (2005) who derived a
disk fraction of $42\%\pm13\%$ and $50\%\pm17\%$ around BDs in
IC\,348 and Chamaeleon\,I respectively. With our model, we
find that disks around BDs in Taurus are all significantly
flaring, 
indicating that heating by the central object is efficient and
that the disks we observe retain a significant amount of gas.
Using H$\alpha$ EW measurements, we find that 6 out of 7 BD with
disks (85\%) are still significantly accreting. 

We find that BDs with disks appear statistically more numerous 
in one of the Taurus filaments, specifically the northern one. As this
filament also contains a very high proportion of CTTs relative
to WTTs,  this segregation could be due to a time evolution
effect if the northern filament is the youngest one, a result
pointing toward a similar formation \& evolution process for
stars and brown dwarfs + disks. However, the age difference
between the different filaments appears  too small to fully
explain such a difference.  Moreover, if the various aggregates
are not exactly in the same evolutionary stage, our result could
imply that the BDs belonging to an older aggregate have lost
their circumstellar disk before the stars, implying that the
disk lifetime depends on the mass of the central object, with
BD disks having a shorter lifetime.  This is in contradiction
to a recent result from Alexander \& Armitage (2006) who
propose that BD disk lifetimes could be larger than the stellar
ones. Alternatively, if all the Taurus aggregates have similar
ages, and if ejection is the only process left to eliminate BD
disks, then we are still to understand why ejection would result
in so different outcomes in aggregates with similar stellar
densities. 

We also compared the underlying $^{13}$CO emission for BDs with
and without disks and found that BDs with disks appear to be
found with stronger molecular emission than BDs without disk,
showing that the presence and/or the size of a disk around a BD
could be linked to the underlying parent core gas density. 
There is no such effect when we compare the BD positions with
the underlying stellar density, but Goodwin et al.\ (2005)
concluded that a lack of difference between stars and brown
dwarfs spatial distribution does not necessarily exclude the
ejection scenario.  Our current study of the spatial
distribution of BDs and their disks does not allow us to
distinguish between the two main BD formation models, although
it provides another piece of evidence that ejection cannot be
the only BD formation process (see e.g., Luhman et al.\ 2006).
Clearly more numerical simulations using initial physical
parameters closely matched to the Taurus aggregates and more
observations to provide unambiguous constraints to these
simulations, are needed to explain the stars and brown dwarfs
spatial distribution as well as the evolution of their
circumstellar environment with time, at a scale significantly
smaller than 10~AU. 

Last but not least, a possible by-product of our study is to
recalibrate the Barrado y Navascu\'es and Mart\'in (2003)
WTTS/CTTS limit in the substellar domain. The global shape of
the limit fits our data points if it is  raised by a factor 1.3.

\begin{acknowledgements}
We thank Cathie Clarke and Anthony Whitworth for enlightening
discussions about BD formation models, and an anonymous referee
for a detailed report that helped clarify many points in our
paper. 
  This research has made use of the CDS database. F.~M. thanks
the ``Center for long Wavelength Astrophysics" for supporting a
visit at JPL. We thank the ``Programme National de Physique
Stellaire (PNPS, CNRS/INSU, France) for financial support.
  \end{acknowledgements}

\end{document}